\documentclass[12pt,preprint]{emulateapj}

\shorttitle{SDSS J141624.08$+$134826.7}
\shortauthors{Cushing et al.}

\newcommand\teff{\mbox{$T_\mathrm{eff}$}}
\newcommand\fsed{\mbox{$f_\mathrm{sed}$}}
\newcommand\logg{\mbox{$\log g$}}

\newcommand\kzz{\mbox{$K_\mathrm{zz}$}}
\newcommand\obj{\mbox{SDSS J1416$+$13A}}
\newcommand\mobj{\mbox{2MASS J1126$-$50}}
\newcommand\tobj{\mbox{SDSS J1416$+$13B}}

\begin{document}

%% LaTeX will automatically break titles if they run longer than
%% one line. However, you may use \\ to force a line break if
%% you desire.

\title{SDSS J141624.08+134826.7: Blue L Dwarfs and Non-Equilibrium
  Chemistry}

%% Use \author, \affil, and the \and command to format
%% author and affiliation information.
%% Note that \email has replaced the old \authoremail command
%% from AASTeX v4.0. You can use \email to mark an email address
%% anywhere in the paper, not just in the front matter.
%% As in the title, use \\ to force line breaks.

\author{Michael C. Cushing\altaffilmark{1}} 
\affil{Jet Propulsion Laboratory, California
  Institute of Technology, MS 264-723, 4800 Oak Grove Drive, Pasadena, CA 91109}
\email{michael.cushing@gmail.com}

\author{D. Saumon}
\affil{Los Alamos National Laboratory, Applied Physics
        Division, MS F663, Los Alamos, NM 87545, dsaumon@lanl.gov}

\author{Mark S. Marley}
\affil{NASA Ames Research Center, MS 254-3, Moffett
        Field, CA 94035, Mark.S.Marley@NASA.gov}

      \altaffiltext{1}{Visiting Astronomer at the Infrared Telescope
        Facility, which i s operated by the University of Hawai`i under
        cooperative Agreement no. NCC 5-538 with the National
        Aeronautics and Space Administration, Office of Space Science,
        Planetary Astronomy Program.}

%\author{Copyright 2010.  All rights reserved.}

\begin{abstract}

  We present an analysis of the recently discovered blue L dwarf SDSS
  J141624.08+134826.7.  We extend the spectral coverage of its published
  spectrum to $\sim$4 $\mu$m by obtaining a low-resolution $L$ band
  spectrum with SpeX on the NASA IRTF.  The spectrum exhibits a
  tentative weak CH$_4$ absorption feature at 3.3 $\mu$m but is
  otherwise featureless.  We derive the atmospheric parameters of SDSS
  J141624.08+134826.7 by comparing its 0.7$-$4.0 $\mu$m spectrum to the
  atmospheric models of Marley and Saumon which include the effects of
  both condensate cloud formation and non-equilibrium chemistry due to
  vertical mixing and find the best fitting model has \teff=1700 K,
  \logg=5.5 [cm s$^{-2}$], \fsed=4, and \kzz=10$^4$ cm$^2$ s$^{-1}$.
  The derived effective temperature is significantly cooler than
  previously estimated but we confirm the suggestion by
  \citeauthor{2010ApJ...710...45B} that the peculiar spectrum of SDSS
  J141624.08+134826.7 is primarily a result of thin condensate clouds.
  In addition, we find strong evidence of vertical mixing in the
  atmosphere of SDSS J141624.08+134826.7 based on the absence of the
  deep 3.3 $\mu$m CH$_4$ absorption band predicted by models computed in
  chemical equilibrium.  This result suggests that observations of blue
  L dwarfs are an appealing way to quantitatively estimate the vigor of
  mixing in the atmospheres of L dwarfs because of the dramatic impact
  such mixing has on the strength of the 3.3 $\mu$m CH$_4$ band in the
  emergent spectra of L dwarfs with thin condensate clouds.

\end{abstract}

\section{Introduction} \label{sec:Introduction}

Field L dwarfs \citep{1999ApJ...519..802K,2005ARA&A..43..195K} comprise
a mix of very low-mass stars and brown dwarfs.  Although effective
temperature (\teff) is the primary atmospheric parameter that controls
their emergent spectra and thus their spectral type
\citep[e.g.,][]{2008ASPC..384...85K}, secondary parameters such as
surface gravity ($g$) and metallicity [Fe/H], as well as the condensate
clouds properties and photospheric vertical mixing also play an
important role.  This is perhaps best illustrated by the fact that the
$J-K_s$ colors of L dwarfs can vary by up to 1 mag at a given spectral
type
\citep{2002ApJ...564..452L,2004AJ....127.3553K,2006AJ....131.2722C,2008ASPC..384...85K,2009AJ....137....1F,2010AJ....139.1808S}.
Low metallicity explains the outliers with extreme blue $J-K_s$ colors,
but the relative importance of variations in metallicity, surface
gravity, condensate cloud properties, and the vigor of vertical mixing
to the bulk of the variation remains unknown.

\begin{figure*} 
\centerline{\includegraphics[width=6in]{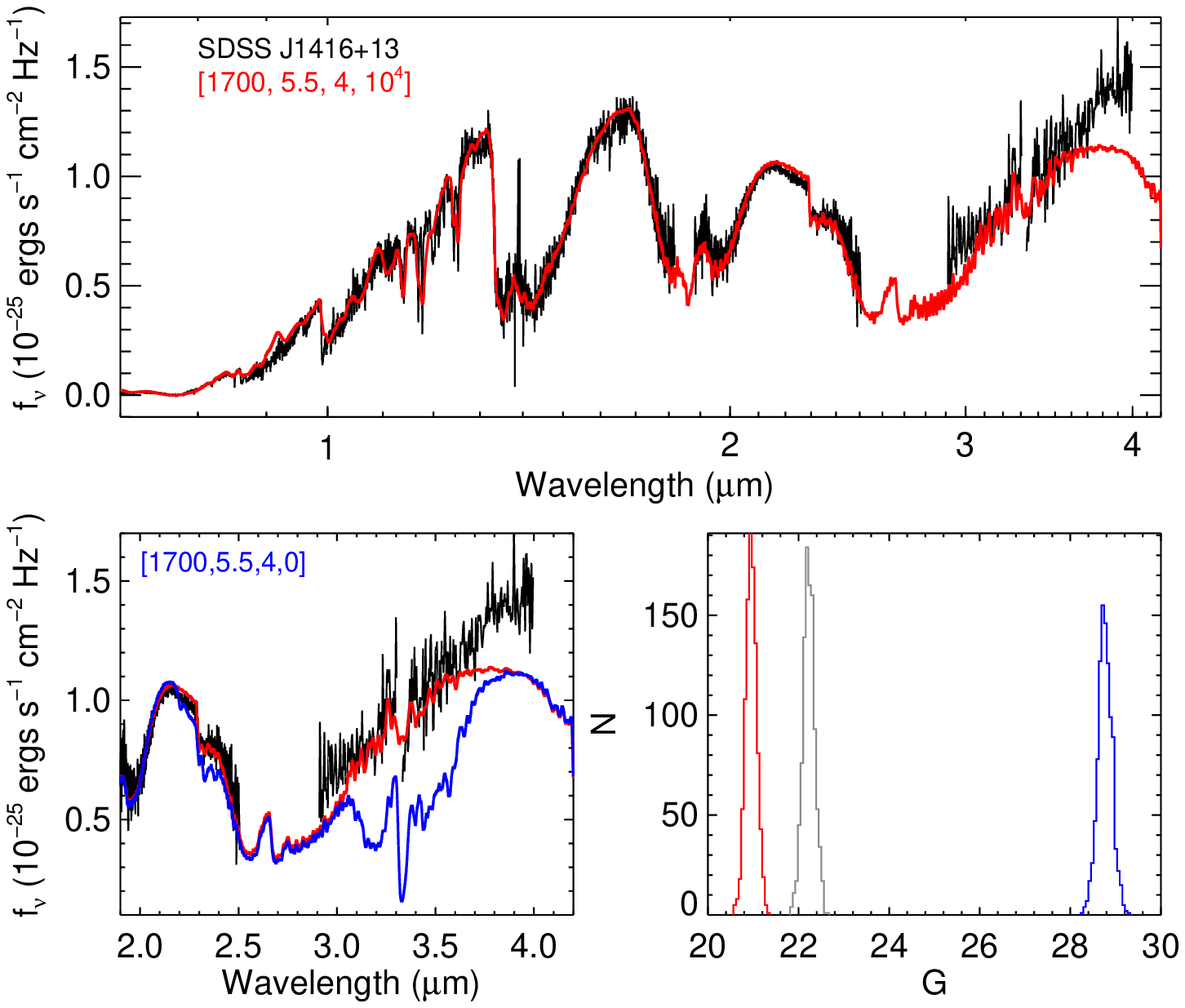}}
\caption{\label{fig:Fit}\textit{Top}: 0.8$-$4.0 $\mu$m spectrum of \obj\
  (black) overplotted with the best fitting model (red) with \teff=1700
  K, \logg=5.5 [cm s$^{-1}$], \fsed=4, \kzz=10$^4$ cm$^2$ s$^{-1}$
  [1700, 5.5, 4, 10$^{4}$].  \textit{Lower left:} Same as top panel
  except the model with \kzz=0 cm$^2$ s$^{-1}$ is shown in blue.  Note
  that for display purposes, the models are shown at a lower resolution.
  \textit{Lower right:} Distribution of $G$ values for the best fitting
  model [1700, 5.5, 4, 10$^4$] in red, the second best-fitting model
  [1800, 5.5, 4, 0] in grey, and the model with \kzz=0 cm$^2$ s$^{-1}$
  (blue) shown in the lower left panel.}
\end{figure*}

L dwarfs that exhibit bluer than average colors also appear to exhibit
peculiar spectral features including enhanced FeH, \ion{K}{1}, and
H$_2$O absorption
\citep{2003AJ....126.2421C,2004AJ....127.3553K,2006AJ....131.2722C,2007AJ....133..439C,2007MNRAS.378..901F,2008ApJ...674..451B}.
All of the secondary parameters described above, along with unresolved
binary, have been invoked to explain these spectral features.
\citet{2008ApJ...674..451B} performed a detailed analysis of the
red-optical and near-infrared spectra of the blue L dwarf 2MASS
J11263991$-$5003550 (hereafter \mobj) and found that thin condensate
clouds provided the overall best explanation of its spectral properties.
However, since the properties of the cloud model used in their analysis
are encapsulated in a free parameter, it is still unclear which, if any,
of the secondary parameters is the underlying physical cause of thin
condensate clouds.  More recent work by \citet{2009AJ....137....1F} and
\citet{2010AJ....139.1808S} have shown the blue L dwarfs have kinematics
consistent with old age, suggesting that surface gravity and/or
low-metallicity may account for their blue colors.

A common proper motion system consisting of a blue L dwarf SDSS
J141624.08+134826.7 \cite[hereafter \obj,
][]{2010AJ....139.1045S,2010ApJ...710...45B,2010MNRAS.404.1952B} and a
blue T dwarf ULAS J141623.94$+$134836.3 \citep[hereafter \tobj,
][]{2010A&A...510L...8S,2010MNRAS.tmp..400B} separated by 9$''$ was
recently discovered that may shed some light on the underlying
atmospheric physics of the blue L dwarfs.  Based on its red optical
spectrum, \citet{2010ApJ...710...45B} and \citet{2010AJ....139.1045S}
classify \obj\ as a dwarf (L6 and L5 respectively), while
\citet{2010MNRAS.tmp..400B} classify it as a dwarf/subdwarf (d/sdL7) and
\citet{2010ApJS..190..100K} classify it as a subdwarf (sdL7).
The subdwarf classifications suggest metallicity may play a role in
explaining the properties of the blue L dwarfs but the disparate
spectral types only underscores the difficultly of separating the
effects of gravity, metallicity, and the condensate clouds properties.
Using model atmospheres \citet{2010AJ....139.2448B} found that the
near-infrared spectrum of \tobj\ was consistent with a subsolar
metallicity of [Fe/H] $\leq$ $-$0.3 and high surface gravity of \logg =
5.2$\pm$0.4 [cm s$^{-2}$].  Since we can reasonably assume that \obj\
and \tobj\ are coeval and have the same composition, this analysis
provides the first concrete evidence that metallicity and/or surface
gravity are the secondary parameters controlling the spectra of the blue
L dwarfs.

In this paper, we extended the spectral coverage of the spectrum of
\obj\ to 4 $\mu$m and derive its atmospheric parameters by comparing its
0.7$-$4.0 $\mu$m spectrum to the atmospheric models of Marley and Saumon
\citep{2002ApJ...568..335M,2008ApJ...689.1327S}.  We confirm the
suggestion of \citet{2010ApJ...710...45B} that the spectral
peculiarities of \obj\ are due to a thin condensate cloud and find
strong evidence for vertical mixing in the atmosphere of \obj.

\section{Observations}

A 1.9$-$4.0 $\mu$m spectrum of \obj\ was obtained on 2010 Jan 29 (UT)
using SpeX \citep{2003PASP..115..362R} on the 3 m NASA IRTF.  We used
the Long-Cross-Dispersed mode (LXD) with the 0$\farcs$8 slit to obtain a
spectrum at a resolving power $R\equiv \lambda/ \Delta \lambda \approx$
940.  A series of 15 sec exposures were obtained at two positions along
the 15$''$ slit.  We observed the A0 V star HD 121996 for telluric
correction and flux calibration purposes.  All observations were
conducted at the parallactic angle even though slit losses and spectral
slope variations due to differential atmospheric refraction are minimal
at these wavelengths.  Finally exposures of internal flat field and Ar
arc lamps were obtained for flat fielding and wavelength calibration.

The data were reduced using Spextool, the IDL-based data reduction
package for SpeX \citep{2004PASP..116..362C} using standard procedures.
The raw spectrum of SDSS 1416$+$13 was corrected for telluric correction
and flux calibrated using the observed A0 V star and the technique
described in \citet{2003PASP..115..389V}.  Regions of low
signal-to-noise (S/N) at wavelengths with strong telluric absorption
from 2.5 to 2.9 $\mu$m and centered at 3.3 $\mu$m were removed and the
spectrum was rebinned to 1 pixel per resolution element resulting in a
final S/N of $\sim$15.

To construct a nearly complete 0.5$-$4.0 $\mu$m spectrum of \obj\ we
combined our $L$-band spectrum with previously published red-optical and
near-infrared spectra.  The LXD spectrum was scaled to match the
$K$-band flux level of the \citet{2010AJ....139.1045S} 0.8$-$2.5 $\mu$m
SpeX spectrum ($R$=2000, S/N=50$-$150) and the two spectra were averaged
together.  The red-optical SDSS spectrum was scaled to match the flux
level of the merged spectrum and then averaged with the merged spectrum.
The resulting 0.5$-$4.0 $\mu$m spectrum was absolutely flux calibrated
using published 2MASS photometry and the technique described in
\citet{2009ApJS..185..289R}.

The spectrum, which is shown in the top panel of Figure \ref{fig:Fit},
is consistent with the $L$-band spectra of other mid-type L dwarfs
\citep{2000ApJ...541L..75N,2005ApJ...623.1115C,2009ApJ...702..154S} in
that it is relatively featureless at this resolution and S/N.  There is
a hint of absorption at 3.3 $\mu$m due to the Q branch of the $\nu_3$
band of CH$_4$ but the telluric CH$_4$ absorption makes this
identification tentative.

We also computed the bolometric flux of \obj\ using the flux calibrated
spectrum.  We first extended the spectrum blueward by extrapolating from
the end of the SDSS spectrum at $\sim$4000 \AA\ to zero flux at zero
wavelength.  Gaps in the spectrum were linearly interpolated over and we
used the \textit{Spitzer} Infrared Camera (IRAC) [4.5] photometry
\citep{2010MNRAS.404.1952B} and a Rayleigh-Jeans tail to account for the
flux emitted at $\lambda > 4$ $\mu$m.  Integrating over the spectrum
yields $f_\mathrm{bol}$=2.13$\pm$0.04 W m$^{-2}$
($m_\mathrm{bol}$=15.19$\pm$0.02)\footnote{$m_{\mathrm{bol}} = -2.5
  \times \log(f_{\mathrm{bol}}) - 18.988$ assuming $L_{\odot} =
  3.86\times10^{26}$ W and $M_{\mathrm{bol}\odot} = +4.74$.} where the
error is generated via a Monte Carlo simulation that includes the errors
in the individual spectral points, the error in the IRAC magnitude, and
the overall absolute flux calibration error of the spectrum.  If
instead, we use the best fitting model identified in the next section to
account for the flux emitted at $\lambda >$ 4.5 $\mu$m we find
$f_\mathrm{bol}$=2.11$\pm$0.04 W m$^{-2}$ which indicates that the
assumed form of the flux distribution longward of 4.5 $\mu$m is not a
significant source of systematic error.  This bolometric flux value can
be used to compute the bolometric luminosity of \obj\ when a precise
parallax becomes available.

\section{Analysis:  Atmospheric Parameters}

We compared the spectrum of \obj\ to the model spectra of
\citet{2002ApJ...568..335M} and \citet{2008ApJ...689.1327S} in order to
estimate its atmospheric parameters (see \citet{2008ApJ...678.1372C} and
\citet{2009ApJ...702..154S} for a more detailed description of the
models).  We used a grid of solar
metallicity\footnote{\citet{2010AJ....139.2448B} found that \tobj\ was
  slightly metal poor at [Fe/H]=$-$0.3 and the dwarf/subdwarf and
  subdwarf spectral types of \citet{2010MNRAS.tmp..400B} and Kirkpatrick
  et al. (submitted) suggest \obj\ has a subsolar metallicity.  However
  nonsolar metallicity cloudy models are not currently available so we
  proceed with our analysis using only solar metallicity models.} models
with \teff=1000 to 2400 K in steps of 100 K, \logg=4.0, 4.5, 5.0, 5.5
[cm s$^{-2}$], \fsed=1,2,3,4, $\infty$ (no cloud), and \kzz=0,10$^4$
cm$^2$ s$^{-1}$.  The sedimentation efficiency \fsed\
\citep{2001ApJ...556..872A} parameterizes the efficiency of condensate
sedimentation relative to turbulent mixing.  Clouds with larger values
of \fsed\ have larger modal particle sizes and thus are thinner.  The
eddy diffusion coefficient \kzz\ parameterizes the vigor of mixing in
the radiative layers of the atmosphere and ranges from 10$^2$ to 10$^5$
cm$^2$ s$^{-1}$ in the stratospheres of giant planets
\citep{2006ApJ...647..552S}.  The model spectra were smoothed to the
spectral resolution of the data and interpolated onto the wavelength
scale of the data.

%Preliminary nonsolar metallicity models with [M/H]=$\pm$0.2 dex
%  produces variations in \teff, \logg, and \fsed\ of $\pm$30 K, $\pm$0.3
%  dex, and $\pm$0.3, respectively \citep{2009ApJ...702..154S}

We identified the best fitting model spectra in the grid using the
goodness-of-fit statistic $G_k$ described in \citet[][see
also \cite{2009ApJ...706.1114B,2008ApJ...689L..53B}]{2008ApJ...678.1372C}.  We
weighted each spectral point by its width in logarithmic wavelengths
$w_i=\delta \ln \lambda_i$, but note that the results do not change if
we give equal weights to all points ($w_i=1$).  For each model $k$, we
compute the scale factor $C_k$=$(R/d)^2$, where $R$ and $d$ are the
radius and distance of the dwarf respectively, that minimizes $G_k$.
The best fitting model is identified as having the global minimum $G_k$
value.  To estimate the uncertainty, we run a Monte Carlo simulation
using both the uncertainties in the individual spectral points and the
overall absolute flux calibration of the spectrum
\citep[see][]{2008ApJ...678.1372C,2009ApJ...706.1114B,2010AJ....139.2448B}.

\begin{figure*} 
\vspace{-2in}
\centerline{\includegraphics[width=6in]{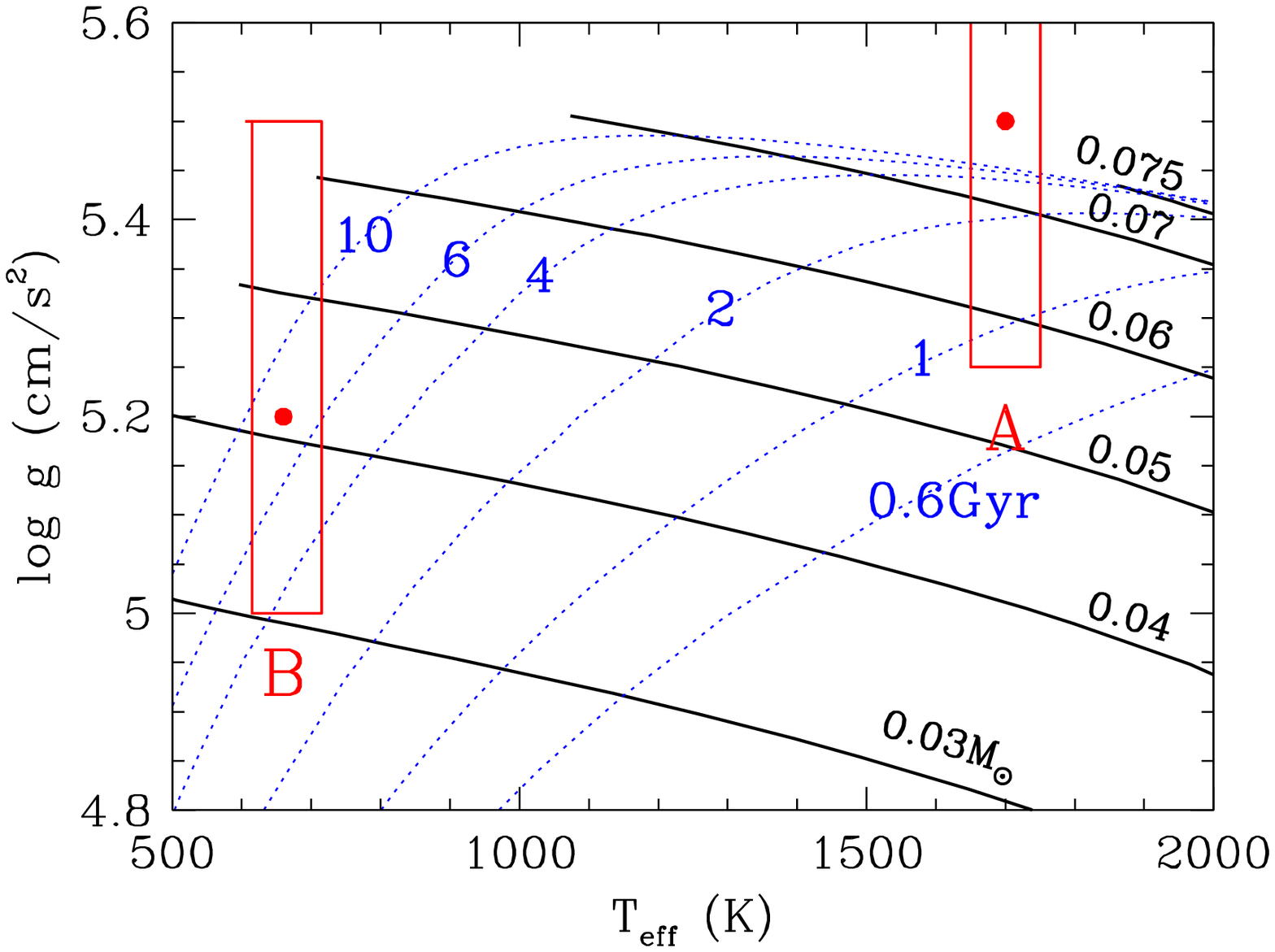}}
\vspace{-1.8in}
\caption{\label{fig:Evo}Age and masses of the two components of \obj\
  based on the (\teff,$g$) derived from the analysis of their spectra
  and an evolution sequence using cloudless atmospheres and [M/H]=$-$0.3
  \citep{2008ApJ...689.1327S}.  The parameters of \obj\,
  (\teff=1700$\pm$50\,K, \logg =5.5$\pm$0.25 [cm s$^{-2}$]) are from
  this work, those of SDSS 1416+13B (\teff=660$^{+55}_{-45}\,$K,
  \logg=5.2$^{+0.3}_{-0.2}$ [cm s$^{-2}$]) are from
  \citet{2010AJ....139.2448B}.  Isochrones are shown with blue dotted
  lines and are labeled with the age in Gyr.  Black solid lines show the
  cooling for brown dwarfs with masses given in $\,M_\odot$.  Isochrones
  with ages greater than 4 Gyr are consistent with the parameters
  derived from both objects.}
\end{figure*}

The best fitting model has \teff=1700 K, \logg = 5.5 [cm s$^{-2}$],
\fsed=4, and \kzz=10$^4$ cm$^2$ s$^{-1}$, (hereafter [1700, 5.5, 4,
10$^4$]), and is shown in the top panel of Figure \ref{fig:Fit}
overplotted on the spectrum of \obj.  The distribution of $G$ values
generated during the Monte Carlo simulation for the best-fitting model
[1700, 5.5, 4, 10$^{4}$] and second best-fitting model with [1800, 5.5,
4, 0] are given in the lower right panel of Figure \ref{fig:Fit} and
indicate that the best-fitting model is 19 $\sigma$ better than the
second best-fitting model.  Overall the model spectrum matches the data
well in the 0.5$-$2.5 $\mu$m wavelength range but it fails to match the
shape of the $L$-band.  In particular, the model turns over at 3.8
$\mu$m while the data continue to rise to the limit of the observations
at $\sim$4 $\mu$m.  This mismatch appears to be systemic
\citep[e.g.,][]{2008ApJ...678.1372C,2009ApJ...702..154S} and is probably
a result of the cloud model not producing enough small ($\le$1 $\mu$m)
particles, but a nonsolar metallicity cannot be ruled out.

The derived effective temperature of \teff=1700 K and surface gravity of
\logg=5.5 [cm s$^{-2}$] are consistent with the values derived for other
L dwarfs with similar spectral types
\citep{2008ApJ...678.1372C,2009ApJ...702..154S,2009A&A...503..639T}.
The effective temperature of 1700 K is, however, substantially cooler
than the 2200 K temperature derived by \citet{2010ApJ...710...45B} using
the AMES-dusty model atmospheres \citep{2001ApJ...556..357A}.  The
AMES-dusty models assume that the dust grains form in chemical
equilibrium with the gas and do not gravitational settle into clouds.
As a result, AMES-dusty models with \teff=1700 K have extremely red
near-infrared colors because of the large column of dust above the
photosphere.  An AMES-dusty model with a higher effective temperature
(i.e., less dust) is therefore required to match the blue colors of
\obj.

The derived surface gravity of \logg=5.5 [cm s$^{-2}$] is not only at
the edge of our model grid but is also somewhat high because at
\teff=1700 K evolutionary models restrict the surface gravities of brown
dwarfs to be $<$ 5.42 in the case of cloudless atmospheres and $<$5.37
in the case of cloudy (\fsed=2) atmospheres \citep{2008ApJ...689.1327S}.
Nevertheless we continue to report results on the grid points because
the model atmospheres were computed at these values.  The high surface
gravity supports the suggestion of \citet{2010AJ....139.2448B} that old
age may be the underlying cause for the thin condensate clouds in the
blue L dwarfs; however, the surface gravities of isolated field L dwarfs
are notoriously difficult to measure precisely \citep[$\pm$ 0.25
dex,][]{2008ApJ...678.1372C} rendering any connection tentative at best.

The derived sedimentation efficiency of \fsed=4 is high for an L dwarf
as values of 1$-$3 are typical \citep{2009ApJ...702..154S}.  Indeed only
one other L dwarf that has been fit with the Marley and Saumon models
has \fsed=4: the blue L dwarf \mobj\ \citep{2008ApJ...674..451B}
discussed in \S1.  Unfortunately, the properties of the model condensate
clouds are encapsulated in the free parameter \fsed\ which is set
independently of the fundamental parameters $g$ and [Fe/H] so no
conclusion can be drawn regarding the underlying physical mechanism for
the thin clouds.  However as noted in \S1, the analysis of the T dwarf
companion by \citet{2010AJ....139.2448B} suggests that the thin
condensate clouds may be a result of the old age and/or metal poor
atmosphere of \obj.  More theoretical work will be required before the
cloud properties can be tied directly to the fundamental properties such
as gravity and metallicity
\citep{2001A&A...376..194H,2003A&A...399..297W,2004A&A...414..335W,2004A&A...423..657H,2006A&A...455..325H}.

The derived eddy diffusion coefficient of \kzz=10$^4$ cm$^2$ s$^{-1}$ is
also at the edge of our model grid but does indicate the presence of
vertical mixing in the atmosphere \obj.  Although it is not surprising
that such mixing is important in shaping the spectra of L dwarfs
\citep{2003IAUS..211..345S,2006ApJ...647..552S,2007ApJ...656.1136S,2007ApJ...655.1079L,2007ApJ...669.1248H,2009ApJ...695.1517L,2009ApJ...702..154S,2009ApJ...695..844G,2010ApJ...710.1627L},
the impact that the non-equilibrium carbon chemistry has on the 3 $\mu$m
spectral region is particularly dramatic in the case of \obj.  The lower
left panel of Figure \ref{fig:Fit} shows the the $K$ and $L$-band
spectrum of \obj, the best fitting model with [1700, 5.5, 4, 10$^4$],
and a model with the same \teff, \logg, and \fsed, but with \kzz=0
cm$^2$ s$^{-1}$.  The model spectrum computed in chemical equilibrium
exhibits a deep CH$_4$ band at 3.3 $\mu$m and a weak CH$_4$ band at 2.2
$\mu$m.

As a consistency check on the derived parameters, we estimated a
spectroscopic parallax for \obj\ using the derived scale factor $C_k$
\citep{2009ApJ...706.1114B}.  We find that $d/R$=11.9$\pm$0.07 pc
$R_\mathrm{Jup}^{-1}$ for the model with [1700, 5.5, 4, 10$^4$],
assuming the (equatorial) radius of Jupiter at 1 bar is 71492 km
\citep{1981JGR....86.8721L}.  The cloudless evolutionary models of
\citet{2008ApJ...689.1327S} give $R$=0.81 $R_\mathrm{Jup}$ for
\teff=1700 K and \logg=5.5 [cm s$^{-2}$] resulting in a distance
estimate of 9.7$\pm$0.1 pc\footnote{The error is derived solely from the
  error in $C_k$ and therefore does not include any errors in \teff\ and
  \logg.}.  The distance estimate is in good agreement with the
parallactic (9.3$\pm$3.0 pc) and spectrophotometric (8.4$\pm$1.9 pc)
distance estimates of \citet{2010ApJ...710...45B} but lies just at the
1$\sigma$ limits of the \citet{2010A&A...510L...8S} parallactic distance
estimate of 7.9$\pm$1.7 pc and the \citet{2010A&A...510L...8S}
spectrophotometric distance estimate of 8.0$\pm$1.6 pc.  The good
agreement between our derived value and previous results implies that
our derived atmospheric parameters are reasonable.

\section{Discussion}

As noted in \S1, \obj\ is a companion of the blue T dwarf \tobj\
\citep{2010A&A...510L...8S,2010MNRAS.404.1952B} and thus it is
reasonable to assume that the objects have a similar age and
composition.  Although we cannot test whether the metallicities of the
two dwarfs are consistent, we can test that their ages inferred from
evolutionary models are consistent.  Figure \ref{fig:Evo} shows the
cloudless evolutionary models with [M/H]=$-$0.3 from
\citet{2008ApJ...689.1327S} along with the range of \teff\, and \logg\,
values derived for \obj\ (this work) and \tobj\,
\citep{2010AJ....139.2448B}.  The uncertainty in the derived surface
gravities translate into a large range of ages, $\tau$ $>$ 0.8 Gyr in
the case of \obj\ and $\tau$ $>$ 3.2 Gyr in the case of \tobj, but the
derived ages are consistent with \obj\ and \tobj\ being coeval.  We note
that using solar metallicity evolutionary models does not significantly
change this conclusion.

\begin{figure} 
\includegraphics[width=3.5in]{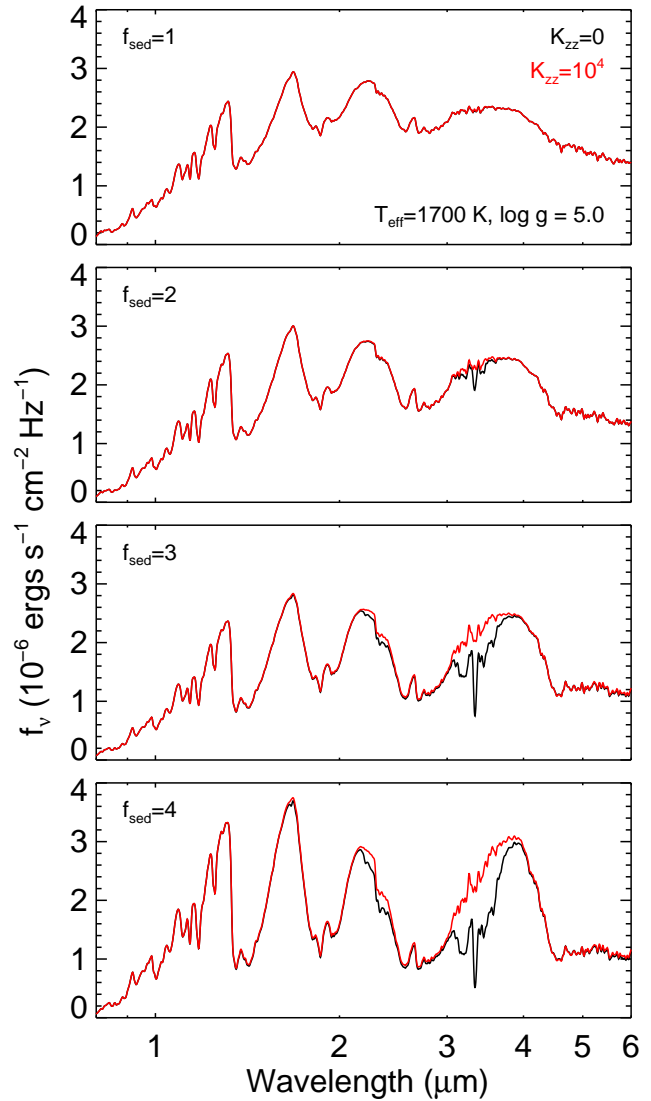}
\caption{\label{fig:ModSeq}Comparison of the effect of non-equilibrium
  chemistry on the band strength of the fundamental CH$_4$ band at 3.3
  $\mu$m as a function of the sedimentation efficiency parameter \fsed.
  Each panel shows two model spectra with \teff=1700 K, \logg=5.0 [cm
  s$^{-2}$], [M/H] = 0, and \kzz=0 (\textit{black}), 10$^4$
  (\textit{red}) cm$^2$ s$^{-1}$ at a fixed \fsed.  The flux density
  units correspond to the emergent flux at the top of the atmosphere.}
\end{figure}

Finally, although there is ample evidence that vertical mixing is
important in the atmospheres of both L and T dwarfs, deriving precise
values of \kzz\ has proven more difficult.  Non-equilibrium chemistry
affects both the carbon and nitrogen chemistry and thus the depths of
the CH$_4$ bands at 2.2 and 3.3 $\mu$m (and to a lesser extent at 7.8
$\mu$m), the fundamental CO band at 4.7 $\mu$m, and the fundamental
NH$_3$ band at 10.5 $\mu$m \citep{2003IAUS..211..345S}.  However not all
of these bands are sensitive to variations in the eddy diffusion
coefficient \kzz.  The NH$_3$ abundance is quenched in convective layers
of the atmosphere rendering the 10.5 $\mu$m band insensitive to
variations of \kzz\ in the radiative layers \citep{2006ApJ...647..552S}.
The fundamental CO band is sensitive to variations in \kzz\ but only
four ground-based spectroscopic observations of T dwarfs
\citep{1997ApJ...489L..87N,1998ApJ...502..932O,2009ApJ...695..844G} at
these wavelengths have been made due to the difficulty observing with
such a high thermal background.  In the absence of spectroscopy,
\textit{Spitzer} photometry of L and T dwarfs at 4.5 $\mu$m with the
Infrared Camera \citep{2004ApJS..154...10F} have provided some
constraints on the vigor of atmospheric mixing \citep[e.g.,
][]{2007ApJ...655.1079L}.

In contrast, $L$ band spectroscopy is relatively easy to obtain from the
ground making the 3.3 $\mu$m region an attractive alternative to the 4.7
$\mu$m CO band.  Figure \ref{fig:ModSeq} shows a sequence of model
spectra with \teff=1700 K, log g=5.0 [cm s$^{-2}$], \fsed=1,2,3,4 and
\kzz=0, 10$^4$ cm$^2$ s$^{-1}$.  The 2$-$4 $\mu$m spectral region forms
above the cloud layer so as the condensate clouds becomes thinner
(\fsed\ increases), these layers become cooler, resulting in an increase
both the abundance and band strengths of CH$_4$.  However, when the
effects of vertical mixing are included (\kzz $>$ 0), the depths of the
CH$_4$ bands are weakened dramatically and are nearly independent of the
cloud properties.  The strong dependence of the strength of the 3.3
$\mu$m CH$_4$ band on $K_{zz}$ in L dwarfs with thin condensate clouds
and the relative ease of obtaining spectra at these wavelengths suggests
that the class of blue L dwarfs are the best objects with which to
constrain the vigor of vertical mixing in the atmospheres of L dwarfs.

\noindent

\acknowledgements

We thank Sarah Schmidt for providing the NIR SpeX spectrum of SDSS
1416$+$13 and Brendan Bowler and Mike Liu for useful discussions.  This
publication makes use of data from the Two Micron All Sky Survey, which
is a joint project of the University of Massachusetts and the Infrared
Processing and Analysis Center, and funded by the National Aeronautics
and Space Administration and the National Science Foundation, the SIMBAD
database, operated at CDS, Strasbourg, France, NASA's Astrophysics Data
System Bibliographic Services, the M, L, and T dwarf compendium housed
at DwarfArchives.org and maintained by Chris Gelino, Davy Kirkpatrick,
and Adam Burgasser, and the NASA/ IPAC Infrared Science Archive, which
is operated by the Jet Propulsion Laboratory, California Institute of
Technology, under contract with the National Aeronautics and Space
Administration.  Support for the modeling work of D.S. was provided by
NASA through the \textit{Spitzer} Science Center.

{\it Facilities:} \facility{IRTF (SpeX)}

\clearpage

\bibliographystyle{apj}
\bibliography{tmp,ref2}

\clearpage

\clearpage

\clearpage

\clearpage

\end{document}